\documentclass[
    ,final            % use final for the camera ready runs
  ]
  {aipproc}

\layoutstyle{6x9}

\newcommand{\be}{\begin{equation}}
\newcommand{\ee}{\end{equation}}
\newcommand{\bea}{\begin{eqnarray}}
\newcommand{\eea}{\end{eqnarray}}

\newcommand{\gs}{\ensuremath{g_s}} % String coupling constant
\newcommand{\ls}{\ensuremath{l_s}} % String length
\newcommand{\lP}{\ensuremath{l_P}} % Planck length
\def\half{\ensuremath{{1\over 2}}}

\def\p{\partial}

\newcommand{\cN}{{\mathcal{N}}}

\newcommand{\bZ}{{\mathbf{Z}}}

\newcommand{\LP}{\ensuremath{L_P}} % WM2 length

\begin{document}

\title{M, Membranes, and OM\footnote{Talk given by A. G\"uijosa at the X Mexican School of
Particles and Fields, Playa del Carmen, M\'exico, November 2002.}}

\author{J.~Antonio Garc\'{\i}a}{
  address={Departamento de F\'{\i}sica de Altas Energ\'{\i}as,
           Instituto de Ciencias Nucleares \\ 
           Universidad Nacional Aut\'onoma de M\'exico,  
           Apdo. Postal 70-543, 
           M\'exico, D.F. 04510}}
\author{Alberto G\"uijosa}{
  address={Departamento de F\'{\i}sica de Altas Energ\'{\i}as,
           Instituto de Ciencias Nucleares \\ 
           Universidad Nacional Aut\'onoma de M\'exico,  
           Apdo. Postal 70-543, 
           M\'exico, D.F. 04510}}
\author{J.~David Vergara}{
  address={Departamento de F\'{\i}sica de Altas Energ\'{\i}as,
           Instituto de Ciencias Nucleares \\ 
           Universidad Nacional Aut\'onoma de M\'exico,  
           Apdo. Postal 70-543, 
           M\'exico, D.F. 04510}}
%email?

\begin{abstract}
We examine the extent to which the action for the membrane of M-theory 
(the eleven-dimensional construct which underlies and unifies all of 
the known string theories) simplifies in the so-called Open Membrane 
(OM) limit, a limit which lies at the root of the various 
manifestations of noncommutativity in the string context.  In order 
for the discussion to be relatively self-contained, we start out by 
reviewing why the strings of ten-dimensional string theory are in fact 
membranes (M2-branes) living in eleven dimensions.  
After that, we recall the definition of OM theory,  
as well as the arguments showing that it is part of
a larger, eleven-dimensional structure 
known as Galilean or Wrapped M2-brane (WM2) theory.
WM2 theory is a rich theoretical 
construct which is interesting for several reasons, in particular 
because it is essentially a toy model of M-theory.  We then proceed to 
deduce a membrane action for OM/WM2 theory, and spell out its 
implications for the four different types of M2-branes one can 
consider in this setting.  
For two of these types, the action in question can be 
simplified by gauge-fixing to a form which implies a \emph{discrete} 
membrane spectrum.  The boundary conditions for the remaining two 
cases do not allow this same gauge choice, and so their dynamics 
remain to be unraveled.
\end{abstract}

\maketitle

%%%%%%%%%%%%%%%%%%%%%%%%%%%%%%%%%%%%%%%%%%%%
%% MAINMATTER
%%%%%%%%%%%%%%%%%%%%%%%%%%%%%%%%%%%%%%%%%%%%
 
\section{From Strings to Membranes}
\label{introsec}

String theory replaces point particles with strings, one-dimensional 
objects whose tension $T_F\equiv 1/2\pi\ls^2$ defines a dimensionful 
parameter $\ls$, known as the string length (which in conventional 
models is expected to be of order the Planck length, $l_s\sim 
10^{-32}$ cm).  Besides moving as a whole, a closed string can 
oscillate in different ways, and upon quantization these internal 
modes give rise to a perturbative spectrum consisting of an infinite 
tower of states with masses $m=2\sqrt{n}/\ls$, $n=0,1,\ldots$ For the 
specific string theory known as Type IIA, the massless states at the 
bottom of the tower correspond to fluctuations of a metric field 
$g_{\mu\nu}$, a rank-2 antisymmetric tensor gauge field $B_{\mu\nu}$, 
a scalar field $\varphi$ (the dilaton) whose vacuum expectation value 
determines the string coupling constant $\gs=\exp\varphi$, a vector 
field $C_{\mu}$, a rank-3 antisymmetric tensor $C_{\mu\nu\lambda}$, 
and the corresponding superpartners.  At low energies ($E\ll 
\ls^{-1}$) these are the only relevant modes; the effective 
field-theoretic description that they provide is known as 
ten-dimensional Type IIA supergravity, with Newton's constant 
$G_N\sim\gs^2\ls^8$.

The non-perturbative spectrum of string theory includes objects known 
as D-branes \cite{polchrr}.  A D$p$-brane is a solitonic 
object extended along $p$ spatial dimensions, 
whose tension (mass per unit $p$-volume) 
is inversely proportional to $\gs$.  Its excitations are 
described by open strings whose endpoints are constrained to lie on 
the brane; quantization of these strings gives rise to another 
infinite tower of states, at the bottom of which there are massless 
states, including a vector gauge field. The effective low-energy
description is consequently in terms of a $(p+1)$-dimensional
supersymmetric gauge theory. 

Type IIA string theory 
contains D-branes with even $p$: a D0-brane (a particle), a D2-brane (a 
membrane), and so on, which are respectively charged under the gauge 
fields $C_{\mu}$, $C_{\mu\nu\lambda}$, etc.  
The general principle at work here is that, just like a point particle
(e.g., the electron) couples to an ordinary vector gauge field ($A_{\mu}$),
an object extended in $p$ spatial
dimensions naturally couples to a rank-$(p+1)$ antisymmetric
tensor gauge field. In particular, the string itself is 
charged under $B_{\mu\nu}$.

Consider a D0-brane.  It has a mass $m=1/\gs\ls$ (known exactly 
because of supersymmetry), and so is very heavy for $\gs\ll 1$, as 
expected for a solitonic object.  On the other hand, for $\gs\gg 1$ 
the D0 mass is the smallest energy scale in the theory.  The dynamics 
of D-particles are such that $n$ D0-branes can form a bound state, 
with mass $m_{n}=n/\gs\ls$.  This evenly-spaced tower of states gives 
rise to a continuum as $\gs\to\infty$, a phenomenon which is 
reminiscent of Kaluza-Klein compactificaction.  Indeed, if we consider 
an \emph{eleven}-dimensional theory in which the $x^{10}$ direction is 
a circle of radius $R_{10}$, we know that the corresponding momentum 
must be quantized, $p_{10}=n/R_{10}$.  A massless eleven-dimensional 
field can thus be expanded in a Fourier series,
\be
\phi(x^{\mu},x^{10})=\sum_{n}e^{inx^{10}/R_{10}}\phi_{n}(x^{\mu}),
\ee
giving rise to an infinite tower of ten-dimensional fields $\phi_{n}$,
with masses $m_{n}^{2}\equiv p_{\mu}p^{\mu}=(n/R_{10})^{2}$.
This would precisely match the D0-brane bound state spectrum if it 
turned out to be the case that
\be \label{rten}
R_{10}=\gs\ls.
\ee

The above agreement is, in fact, more than a coincidence.  
Ten-dimensional Type IIA supergravity, which as mentioned before is 
the low-energy approximation to Type IIA string theory, has been known 
for many years to be directly related to supergravity in \emph{eleven} 
dimensions, with the additional dimension a circle of radius $R_{10}$.  
More precisely, Type IIA supergravity can be obtained by restricting 
the fields of eleven-dimensional supergravity (a metric $g_{MN}$, a 
rank-3 gauge field $A_{MNP}$, and a gravitino $\Psi^{M}_{\alpha}$) 
to be constant along $x^{10}$, i.e., truncating their Kaluza-Klein 
expansions down to the $p_{10}=0$ modes.  When the circle is small, 
these modes have masses much lower than the rest, and so the 
truncation in question (known as dimensional reduction) is physically 
justified.  It is then natural to wonder whether this correspondence 
at the level of Type IIA supergravity could extend somehow to the 
regime where $R_{10}$ is not small, where it would necessarily have to 
involve the $p_{10}=n/R_{10}\neq 0$ modes.

The answer to this question lies in the precise form of the mapping 
between the various supergravity fields.  In particular, the 
$\mu$-$10$ component of the eleven-dimensional metric, which from the 
ten-dimensional perspective is the gauge field that couples to the 
Kaluza-Klein charge $n=p_{10}R_{10}$, correponds in Type IIA language 
to the gauge field $C_{\mu}$, which as we saw before, couples to 
D0-brane charge.  So the D-particle bound states present in Type IIA 
string theory have precisely the right properties to match the full 
Kaluza-Klein tower of eleven dimensional supergravity.  Moreover, the 
$10$-$10$ component of the metric, which controls the size of the 
eleventh dimension and is a scalar from the ten-dimensional perspective, 
translates into the dilaton field $\varphi$, which determines the 
string coupling constant.  The precise relation is in fact 
(\ref{rten}).  Finally, the Type IIA field $B_{\mu\nu}$, which couples 
to the fundamental string, descends from the eleven-dimensional 
rank-3 tensor gauge field $A_{\mu\nu\{10\}}$, which would naturally
couple to a \emph{membrane}.

The conclusion then is that Type IIA string theory is secretly 
eleven-dimensional, and its fundamental degree of freedom, the string, 
is in fact a membrane (the `M2-brane') wrapped around the hidden 
dimension \cite{m}.  The well-known connection at 
the level of supergravity extends to the full string theory, which is 
understood then to be a special (small $R_{10}$) limit of an 
eleven-dimensional theory.  This larger theory has been provisionally 
baptized M-theory (with `mystery' one of the intended meanings); from 
the preceding discussion we know that eleven-dimensional supergravity 
gives its effective low-energy description.  
The situation is summarized in Fig.~1.
M-theory can be shown to 
englobe not only Type IIA but also all of the other known string 
theories, which are thus understood to be part of a single unified 
framework.

%%%%%Fig with 1+1 NCOS
\begin{figure}[bt]
\begin{picture}(150,40) {\small
\put(13,37){\begin{minipage}[t]{6cm} \begin{center}
$9+1$ IIA String Theory
\end{center}\end{minipage}}
\put(13,10){\begin{minipage}[t]{6cm} \begin{center}
$9+1$ IIA Supergravity 
\end{center}\end{minipage}}
\put(83,37){\begin{minipage}[t]{5.3cm} \begin{center}
$10+1$ M-theory
\end{center}\end{minipage}}
\put(83,10){\begin{minipage}[t]{6cm} \begin{center}
$10+1$ Supergravity
\end{center}\end{minipage}}
\put(66,37){\vector(+1,0){28}}
\put(74,40){$\gs\to\infty$}
\put(92,10){\vector(-1,0){28}}
\put(74,13){$R_{10}\to 0$}
\put(31,33){\vector(0,-1){16}}
\put(34,25){$E\ll\ls^{-1}$} }
\put(103,33){\vector(0,-1){16}}
\put(105,25){$E\ll\lP^{-1}=\gs^{-1/3}\ls^{-1}$}
\end{picture}
\vspace*{-0.14cm}
\caption{\small Summary of the connections between
the ten- and eleven-dimensional theories discussed in
the main text. 
The most important point is that the strong-coupling
limit of ten-dimensional Type IIA string theory is in fact 
a theory in \emph{eleven} dimensions.}
\end{figure}
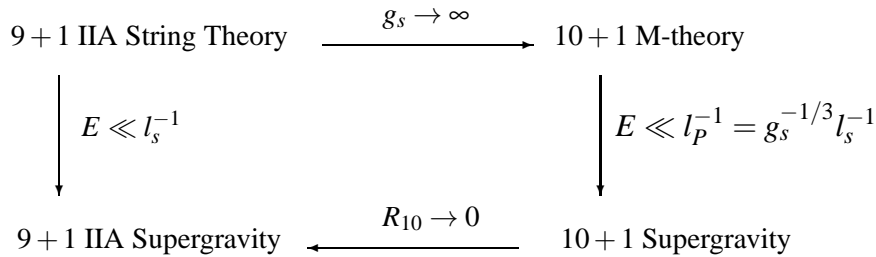

In the effort to understand this mysterious theory, the obvious first 
thing to try is to quantize the M2-brane.  As the membrane moves about 
in eleven-dimensional spacetime, it sweeps out a three-dimensional 
`worldvolume,' which can be described through an embedding function 
$X^{M}(\tau,\sigma,\rho)$.  Unfortunately, the natural (bosonic part of the) 
action for the M2-brane, its tension times the volume it sweeps out,
\be \label{sm2}
S_{\mbox{\footnotesize M2}}=-T_{\mbox{\footnotesize M2}}
\int d\tau d\sigma d\rho\, 
    \sqrt{-\det g_{MN}\p_{\alpha}X^{M}\p_{\beta}X^{N}}~,
\ee
is a complicated non-linear constrained system which has proven extremely  
difficult to quantize.  Essentially all of the progress that has been 
made is based on a discretized version of $S_{\mbox{\footnotesize 
M2}}$ that employs $N\times N$ matrices (the continuous membrane 
being approached in the $N\to\infty$ limit) \cite{whn}.  Quantization 
of this model was found to yield a continuous spectrum \cite{wln}.  
This can be understood at an intuitive level by noting that the action 
(\ref{sm2}) assigns to the membrane an energy proportional to its 
area.  As a consequence, the membrane can develop arbitrarily long 
spikes of infinitesimal area, at zero energy cost.  It is the 
existence of these `flat directions' in the membrane potential 
(together with the supersymmetry-induced cancellation of zero-point 
energies) that gives rise to a continuous spectrum.  This result is in 
sharp contrast with the discrete spectrum of the string, and was 
initially a source of disappointment.

Years later, and following a quite independent line of development, 
the discretized membrane model of \cite{whn} resurfaced (under the 
name Matrix theory) as a proposal for a non-perturbative definition of 
M-theory, restricted to the specific kinematic setup known as the 
infinite momentum frame \cite{bfss}.  In this context, the continuous 
spectrum of the model, previously believed to be a flaw, was 
recognized as a virtue: it is a sign that the membrane yields a 
second-quantized description, with a spectrum that includes 
multiple-particle states. An $n$-particle state is 
obtained by deforming the membrane into $n$ blobs connected by 
infinitesimally thin tubes, which carry no energy.  In this way, a 
single membrane leads to configurations which are indistinguishable 
from multiple-membrane states.

Despite the success of the Matrix proposal, the search is on for new 
ideas which could lead to a less kinematically-restrictive (and 
hopefully more manageable) formulation of M-theory.  In particular, 
the desire to obtain a covariant definition of M-theory naturally 
fuels the ongoing attempts to quantize the membrane covariantly (see 
\cite{nicolai} for some interesting recent developments).  Given the 
complexity of this task, an alternative strategy would be to look for 
an interesting limit in which $S_{\mbox{\footnotesize M2}}$ 
simplifies.  We will describe such a limit in the next section.

\section{OM/WM2 Theory}

Consider a D4-brane (or a stack of them) in Type IIA string theory at 
low energies, $E\ll \ls^{-1}$.  Whereas ordinarily this system would 
have an effective description in terms of a standard (supersymmetric) 
five-dimensional gauge theory, it was discovered in \cite{ncos} 
that, in the presence of a constant background $B_{01}$ field (and 
with appropriately adjusted values of this $B$-field, the metric, and 
$\gs$), the description is in terms of a five-dimensional 
Noncommutative Open String (NCOS) theory (which displays 
noncommutativity between space and time).  What is most remarkable 
about this is that, despite the fact that we are considering a low 
energy regime, by fine-tuning the relevant parameters we manage to 
retain not just the massless modes but the whole infinite tower of 
open string excitations.  On the other hand, the low-energy limit does 
remove from the spectrum the usual closed strings, and in particular 
the graviton.  The result was therefore initially believed to be a 
non-gravitational purely open string theory.

Subsequent work showed that the story is more complicated than that.  
If $x^{1}$ (the direction of the `electric' $B$-field) is not the real 
line, but a circle of radius $R$, then closed strings are in fact 
present in the spectrum of the theory, but only if they wind around 
the circle in the positive direction (i.e., if they have strictly 
positive `winding number,' $w>0$) \cite{km}.  Notwithstanding the fact 
that they coexist with relativistic open strings, these wound closed 
strings obey a non-relativistic dispersion relation, $p_{0}\propto 
p_{\perp}^{2}/2wR + \mbox{oscillators}$.  For finite $R$ these closed 
strings are able to leave the D4-brane(s) and move about freely in 
ten-dimensional spacetime, which allows them to be studied even in the 
absence of the brane(s).  The conclusion is that five-dimensional 
NCOS is actually part of a larger, ten-dimensional string theory, 
which in addition to D4-branes extending along direction $1$ contains 
other objects, including (wound) closed strings and D$p$-branes 
oriented in various ways \cite{wound,go}.  
Gravity also turns out to be present, but 
in a more rudimentary form: it is Newtonian when the theory is 
formulated on a flat background \cite{go,newt}, and `asymptotically 
Newtonian' in a more general background \cite{newt,sahakian2}.  The 
theory in question is thus a drastically simplified version of Type 
IIA string theory; it is known as Type IIA Wound (WIIA) or 
Non-relativistic string theory.

Given the connection between Type IIA string theory and M-theory, it 
is natural to inquire about the eleven-dimensional origin of the above 
setup.  A D4-brane, whose excitations are described by open strings, 
turns out to have
as its M-theoretic counterpart a \emph{fivebrane} with one direction 
wrapped around the $x^{10}$ circle, an object whose excitations are 
described by open M2-branes ending on it.  In addition, we 
know that the $B_{01}$ field included in the NCOS setup descends from 
$A_{01\{10\}}$ in eleven dimensions.  Assembling these facts together, 
five-dimensional NCOS theory is understood to be a special limit of a 
\emph{six}-dimensional theory, which is expected to admit a 
description in terms of open M2-branes terminating on the 
fivebrane(s), and to possess a generalized form of noncommutativity.  
This M-theoretic structure, known as Open Membrane (OM) theory 
\cite{om}, plays a role analogous to that of M theory itself: it 
underlies and unifies all of the noncommutative theories which 
originate from string theory, be they of the open brane 
\cite{ncos,om,odp} or of the purely field-theoretic 
\cite{sw} type.

The embedding of NCOS into Wound string theory implies of course an 
analogous embedding for the OM case.  Indeed, Wound IIA string theory 
can be lifted to eleven dimensions to obtain what is known as Wrapped 
\cite{wound} or Galilean \cite{go} M2-brane (WM2) theory, an 
M-theoretic construct which contains OM theory as a special class of 
states \cite{wound}.  To be precise, OM theory corresponds to those 
states of WM2 theory that involve M5-branes extended along the 
`longitudinal' directions $1$-$10$ (the directions singled out by the 
background $A$ field).  WM2 theory contains in addition (partially or 
fully) transverse M5-branes, closed M2-branes, and Newtonian gravity 
\cite{wound,go,newt}, and includes all Wound string and Wrapped brane 
theories \cite{wound,go} (and consequently all noncommutative open 
brane theories) as special limits.  It is clearly desirable to 
increase our knowledge about this rich theoretical structure, which 
constitutes a simplified model of M theory.  The question for us then 
becomes, what happens to $S_{\mbox{\footnotesize M2}}$ in the OM/WM2 
theory limit?

\section{OM/WM2 Action}

The bosonic part of the action for an M2-brane in a background 
$A_{012}$ field is
\be \label{ngmem}
S_{\mbox{\footnotesize M2}}=-T_{\mbox{\footnotesize M2}}
\int d^3\sigma \left[ \sqrt{-\det g_{MN}\p_{\alpha}X^{M}\p_{\beta}X^{N}}
-A_{012}\varepsilon^{\alpha\beta\gamma}\p_{\alpha}X^{0}
\p_{\beta}X^{1}\p_{\gamma}X^{2}\right]~,
\ee
where $T_{\mbox{\footnotesize M}}=1/(2\pi)^2\lP^3$
(with $\lP$ the eleven-dimensional Planck length) is the membrane 
tension, the worldvolume coordinates 
$\sigma^{\alpha}\equiv(\tau,\sigma,\rho)$, $\varepsilon^{012}=+1$, and 
the spacetime indices $M,N=0,\ldots,10$.  
Notice that, for ease of notation, we have made a slight
change of conventions, relabeling as $x^{2}$ the coordinate
which in the previous sections was denoted $x^{10}$. 
In the following, directions $1$ and $2$ will both be assumed
to be circles, with respective radii $R_{1}$ and $R_{2}$.

In the OM/WM2 limit (and 
after partial gauge-fixing), the action (\ref{ngmem}) can be shown to 
reduce to
\cite{membrane}
\bea \label{nromem}
S_{\mbox{\footnotesize W}}&=&
-T_{\mbox{\footnotesize W}} \int d^{3}\sigma \left[-\half \dot{Y}^2
+\half (X'^2\hat{Y}^2-2X'\cdot\hat{X}Y'\cdot\hat{Y}+\hat{X}^2 
Y'^2) \right. \\
{}&{}&\qquad\qquad\quad +\left. 
l_a(\dot{X}^a-\varepsilon^{abc}X'_b\hat{X}_c)
+\lambda\varepsilon^{\alpha\beta\gamma}
\p_{\alpha}X^0\p_{\beta}X^1\p_{\gamma}X^{2}
\right]~, \nonumber
\eea
where $T_{\mbox{\footnotesize W}}\equiv 1/(2\pi)^2\LP^3$ is the effective membrane
tension, $X^a$ ($a=0,1,2$) and $Y^i$ ($i=3,\ldots,10$) stand respectively 
for the longitudinal and transverse coordinates, $l_a$ are Lagrange 
multipliers enforcing the gauge conditions, and dots, primes and hats 
denote $\tau$-, $\sigma$- and $\rho$-derivatives, respectively.  This 
action must be supplemented with the constraint that the corresponding 
energy-momentum tensor vanish, $T_{\alpha\beta}=0$.  (See 
\cite{membrane} for further details.)
 
The interesting question now is whether it is any easier to quantize 
the WM2 membrane action (\ref{nromem}) than the original action 
(\ref{ngmem}).  The answer turns out to be yes and no \cite{membrane}.  
The first thing to note is that there are in fact several distinct 
cases to consider, depending on whether the membrane is closed or 
open, and if open, whether it ends on (and thus describes
the dynamics of) a longitudinal, partially 
transverse, or fully transverse M5-brane
(e.g., fivebranes
extending along directions 012345, 013456, and 034567, respectively).  
For the closed membrane, 
and for the open membrane associated with a partially 
transverse\footnote{Notice this corrects an erroneous statement in \cite{membrane}.} 
fivebrane, 
the boundary conditions allow one to completely fix the gauge via the 
`static gauge' choice
\be \label{sgauge}
X^0=c\tau, \quad X^1=w_1 R_1\sigma, \quad X^{2}=w_{2} R_{2}\rho,
\ee
with $c$ an arbitrary constant and $w_1,w_{2}\in\bZ$,
thereby reducing (\ref{nromem}) to the free-field action
\be \label{nrsmem}
S_{\mbox{\footnotesize W}}^{\mbox{\footnotesize (s)}}=
-T_{\mbox{\footnotesize W}}\int d^3 x \left[{1\over 2}\p_a Y\cdot \p^a Y
+\lambda\right]~,
\ee
which describes a non-relativistic membrane.
The resulting energy spectrum is \cite{membrane}
\be \label{p0cmem}
p_{0}=\lambda {wR_1 R_{2}\over\LP^3}
+{\LP^3 p_{\perp}^2\over w R_1 R_{2}}
+{\cN\over wR_1 R_{2}}~,
\ee
where $w\equiv w_1 w_{2} >0$ is the membrane wrapping number,
and we have defined a number operator
\be \label{calN}
\cN\equiv\sum_{\vec{n}}\sqrt{(n_1 w_1 R_1)^2+(n_{2} w_{2} R_{2})^2}
a^{\dagger}_{\vec{n}}\cdot a_{\vec{n}}~.
\ee
We thus learn that, in 
contrast with the standard membrane, the spectrum of the closed WM2 
membrane (and that of the open membrane ending on a partially
transverse M5-brane) is \emph{discrete}.  This is of course due to the 
non-relativistic character of $S_{\mbox{\footnotesize 
W}}^{\mbox{\footnotesize (s)}}$: as is evident in (\ref{nrsmem}), the 
membrane potential does not have flat directions.
As a check on the result (\ref{p0cmem})-(\ref{calN}), one 
can verify that, under reduction to ten dimensions, the expected 
NCOS/WIIA spectra \cite{ncos,wound,go}
are correctly reproduced. (In addition, one obtains
an interesting prediction for the `longitudinal NS5-brane'
of WIIA theory.)

The remaining two cases involve an open membrane that ends on either a 
longitudinal or a fully transverse M5-brane.  These cases are 
particularly interesting: the first because it is precisely the OM 
theory setup, the second because the fivebrane in question is 
tensionless \cite{membrane}, and would therefore be expected to play 
an important role in the dynamics of the theory.  Unfortunately, for 
these cases the boundary conditions are incompatible with the choice 
of static gauge, and so the system remains complicated.  One can 
actually show that the potential following from (\ref{nromem}) has 
flat directions: just like (\ref{ngmem}), it assigns zero energy to 
arbitrarily long but infinitesimally thin spikes \cite{membrane}.  Our 
expectation is then that the spectrum of excitations is continuous 
(and in particular includes multi-particle states).

\section{Conclusions}

We have reviewed the story of the passage from strings in
ten dimensions to membranes in eleven dimensions \cite{m}.
The punchline of this story is the existence of 
a mysterious eleven-dimensional structure known as M-theory,
which underlies and unifies all of the known string theories.
We have then argued that OM/WM2 theory \cite{om,wound,go} 
is an interesting simplified version of M-theory, and 
consequently a good setting to try to improve our understanding
of the basic M-theoretic degrees of freedom. 

Our main result in this direction is
the derivation of an explicit membrane action, Eq.~(\ref{nromem}),
for OM/WM2 theory \cite{membrane}.
After gauge-fixing, this action was seen to yield
\emph{discrete} excitation spectra for the closed membrane 
and for the `partially transverse' fivebrane (in
the approach we adopt, the latter is described through open membranes
ending on it).  Upon their reduction to ten dimensions
and their reinterpretation in the
language of the corresponding (NCOS/WIIA) string theory 
\cite{ncos,wound,go}, these spectra
correctly reproduce known results (and yield an interesting
prediction for the `longitudinal NS5-brane' spectrum).
For the `longitudinal' and `fully transverse' fivebranes,
on the other hand, progress is hampered by the more complicated
form of the membrane boundary conditions. 
These two cases are particularly interesting---
the former because it is precisely the OM-theory setup \cite{om}, 
and is consequently directly related to various
noncommutative theories;
the latter because it
involves a fivebrane which is known to be \emph{tensionless}
\cite{membrane}, and is therefore expected to play an important
role in the dynamics of WM2 theory. 
The membrane potential for these cases can be seen
to possess flat directions, suggesting continuous excitation
spectra. These intriguing sytems clearly deserve further study.

\section*{Acknowledgments}
The work of AG was supported by 
Mexico's National Council of Science 
and Technology (CONACyT), under grant I39233.
JAG and JDV 
acknowledge partial support from the grants DGAPA-UNAM 
IN117000 and CONACyT-32431-E.

\vfill\eject
\end{document}